\documentclass[a4paper,11pt]{article}
\usepackage{pos}

\usepackage{bm}
\usepackage{graphicx,xcolor}

\DeclareMathOperator{\tr}{tr}

\usepackage{hyperref}

\title{Direct numerical simulation of the 't~Hooft partition function and (de)confining phases}

\author*[a]{Okuto Morikawa}
\author[b]{Hiroshi Suzuki}

\affiliation[a]{Center for Interdisciplinary Theoretical and Mathematical Sciences
(iTHEMS), RIKEN, Wako 351-0198, Japan}

\affiliation[b]{Department of Physics, Kyushu University, 744 Motooka,
Nishi-ku, Fukuoka 819-0395, Japan}

\emailAdd{okuto.morikawa@riken.jp}

\abstract{%
The 't~Hooft partition function $Z_{\mathrm{tH}}[E_i;B_{ij}]$ is a discrete Fourier transform of Yang--Mills partition functions in background $\mathbb{Z}_N$ 2-form gauge fields and encodes information on confinement, Higgs, Coulomb and oblique-confining phases.
We report a direct Monte Carlo strategy to measure $Z_{\mathrm{tH}}$ without reweighting, by extending hybrid Monte Carlo to include dynamical updates of the background flux variables.
As a first application we measure all flux sectors of four-dimensional $SU(2)$ lattice Yang--Mills on $T^4$ and observe the characteristic ``light/heavy'' behavior expected in the confining phase, together with the shift implied by the Witten effect at $\theta=2\pi$.
We also present a preliminary finite-temperature study and discuss outstanding issues on thermalization and separability between different flux sectors.}

\FullConference{The 42nd International Symposium on Lattice Field Theory (LATTICE2025)\\
2-8 November 2025\\
Tata Institute of Fundamental Research, Mumbai, India\\}

\begin{document}
\maketitle

\section{Motivation and overview}

Generalized global symmetries, and in particular higher-form center symmetries of Yang--Mills (YM) theory, provide a sharp language to characterize phases and to formulate anomaly constraints.
On the lattice, coupling the $\mathbb{Z}_N$ 1-form center symmetry to a background 2-form gauge field $B$ is equivalent to imposing 't~Hooft twisted boundary conditions, labeled by discrete fluxes $z_{\mu\nu}\in\mathbb{Z}_N$.
A crucial consequence is the fractionality of the topological charge in a background $B$, which leads to a mixed 't~Hooft anomaly between the center symmetry and the $2\pi$ periodicity of the $\theta$ angle.
These structures suggest that partition functions in fixed flux sectors, and their discrete Fourier transforms, are natural order parameters for confinement and its variants.

In this proceedings contribution, we summarize results from two companion papers~\cite{Abe:2024fpt, Morikawa:2025ldq} and recent progress.
Our main technical contribution is a ``halfway-updating'' hybrid Monte Carlo (HMC) algorithm that samples both link variables and discrete background flux fields, enabling \textit{direct} estimation of the partition function $Z[B]$ and the 't~Hooft partition function $Z_{\mathrm{tH}}$.
This approach provides information on (de)confinement, Higgs, Coulomb and oblique-confining phases.

\section{Center symmetry, background 2-form fields, and fractional topological charge}

We consider $SU(N)$ lattice YM theory on a four-torus.
The $\mathbb{Z}_N$ 1-form center symmetry acts on link variables $U_\ell\in SU(N)$ by phases determined by intersection numbers with a codimension-2 surface~$\Sigma$.
Coupling to a background 2-form gauge field $B_p\in\mathbb{Z}_N$ modifies plaquette terms schematically as
\begin{equation}
  S[U,B] \sim \sum_{p} \mathrm{Re} \tr \left(e^{-2\pi i B_p/N} U_p\right),
  \label{eq:coupleB}
\end{equation}
and is invariant under the combined transformations of link phases $U\to e^{2\pi i\lambda/N} U$ and $B\to B + d\lambda$.
Globally, this background is encoded by 't~Hooft twisted boundary conditions with fluxes $z_{\mu\nu}=\sum B_{\mu\nu}\bmod N$~\cite{tHooft:1979rtg}.\footnote{In what follows, we assume that the 't~Hooft fluxes $z$ exist, which means that the background 2-form gauge field is flat: $dB=0\bmod N$.}

In the presence of nontrivial flux, the topological charge can become fractional.
We generalize the partition function by multiplying the $\theta$-term $e^{-i\theta Q}$.
For example, for twisted boundary conditions one has~\cite{vanBaal:1982ag}
\begin{equation}
  Q = \frac{1}{16\pi^2}\int \tr F\Tilde F
  = -\frac{\varepsilon_{\mu\nu\rho\sigma} z_{\mu\nu} z_{\rho\sigma}}{8N} + \mathbb{Z}
 \in \frac{1}{N}\mathbb{Z},
  \label{eq:fractionalQ}
\end{equation}
which implies that the partition function in background $B$ is not strictly $2\pi$ periodic in $\theta$ but instead obeys
\begin{equation}
  Z_{\theta+2\pi}[B] = e^{-2\pi i Q[B]} Z_{\theta}[B].
  \label{eq:thetaAnomaly}
\end{equation}
The explicit form (or local description) of~$Q[B]\in\frac{1}{N}\mathbb{Z}$ and the existence of the mixed 't~Hooft anomaly even on the lattice were proved in~Refs.~\cite{Abe:2023ncy} based on Ref.~\cite{Luscher:1981zq}.

This mixed anomaly underlies the Witten effect relation for the 't~Hooft partition function discussed below.

\section{The 't Hooft partition function}

Let $Z[B]$ denote the YM partition function in a fixed background $B_{\mu\nu}\in\mathbb{Z}_N$.
The 't~Hooft partition function is defined as a discrete Fourier transform over temporal flux components,
\begin{equation}
  Z_{\mathrm{tH}}[E_i;B_{ij}]
  \equiv
  \frac{1}{N^3}\sum_{B_{i4}=0}^{N-1}
  \exp\left(\frac{2\pi i}{N}\sum_{i=1}^3 E_i B_{i4}\right) Z[B],
  \label{eq:ZtH}
\end{equation}
where $E_i\in\mathbb{Z}_N$ can be viewed as discrete ``electric'' fluxes and $B_{ij}$ as fixed ``magnetic'' fluxes.
$Z_{\mathrm{tH}}$ is expected to detect distinct quantum phases (confinement/Higgs/Coulomb and oblique variants) and is closely related to the Wilson--'t~Hooft classification of line operators~\cite{Tomboulis:1985ah, Kitano:2017jng, Nguyen:2023fun}.

Traditionally, the free energy differences between flux sectors have been estimated by reweighting~\cite{Kovacs:2000sy, deForcrand:2001nd}, say $Z[z\neq0]/Z[z=0]$ in terms of the fluxes.
Our goal is to measure $Z[B]$ and hence $Z_{\mathrm{tH}}$ \textit{directly}, by constructing a Markov chain whose stationary distribution includes the discrete flux variables.
To do this, for simplicity, we take gauge transformations to set
\begin{align}
 B_{\mu\nu}(x) =
 \begin{cases}
  z_{\mu\nu} & \text{$x_\mu=L-1$ and $x_\nu=L-1$}\\
  0 & \text{otherwise} .
 \end{cases}
\end{align}
This is a representative of an equivalence class, and there are $N^6$ patterns of different classes of~$B$.

\section{Halfway-updating HMC with dynamical $B$ fields}

\subsection{Novel HMC algorithm for $SU(N)/\mathbb{Z}_N$ theory}

We outline the algorithmic idea for the joint sampling of link variables $U$ and discrete background fields $B$.
Starting from a configuration $(U,B)$, we:
\begin{enumerate}
\item Generate conjugate momenta $\pi$ for $U$ from a Gaussian distribution.
\item Evolve $(U,\pi)$ under molecular dynamics (MD) for half a trajectory length $\tau/2$ using the Hamiltonian $H(U,\pi,B)=\pi^2/2 + S[U,B]$, reaching $(\Check U,\Check\pi)$.
\item Update the discrete field by a proposal kernel $P_F(B\to B')$ that is symmetric,
$P_F(B\to B')=P_F(B'\to B)$.
\item Continue the MD evolution for another half step $\tau/2$ using $H(\Check U,\Check\pi,B')$, reaching $(U',\pi')$.
\item Accept/reject the composite move $(U,B)\to(U',B')$ with the Metropolis acceptance probability $\min\{1,\exp(-\Delta H)\}$, where $\Delta H=H(U',\pi',B')-H(U,\pi,B)$.
\end{enumerate}

The symmetry of $P_F$ ensures detailed balance for the joint Boltzmann distribution.
A uniform proposal over flux sectors---$SU(N)/\mathbb{Z}_N$ YM theory with uniformly random $P_F$---is particularly effective in reducing autocorrelation by mixing boundary conditions~\cite{Abe:2024fpt}.

\subsection{Estimating $Z[B]$ and $Z_{\mathrm{tH}}$ by counting}

Because $B$ is part of the dynamical Markov chain, we can estimate the \textit{normalized} weight of each flux sector by counting occurrences:
\begin{equation}
  \widehat Z[B] \equiv \frac{N_{\mathrm{conf}}(B)}{N_{\mathrm{total}}}\,.
  \label{eq:counting}
\end{equation}
This estimator corresponds to $Z[B]/\sum_{B}Z[B]$ (i.e., the probability of sector $B$ in the extended ensemble).
On $T^4$, we additionally average over Euclidean rotations of the flux tuple
\begin{equation}
    (B_{12},B_{13},B_{14},B_{23},B_{24},B_{34})
\end{equation}
when appropriate to improve statistics.
In the $T^4$ study reported in Ref.~\cite{Morikawa:2025ldq}, we used $N_{\mathrm{total}}=2590$ configurations of~$SU(2)$ YM at $\beta=2.6$ and the box size~$L_x\times L_y\times L_z\times L_t=20\times20\times20\times20$.

Having estimated $\widehat Z[B]$ for all relevant $B$, we obtain $Z_{\mathrm{tH}}$ from Eq.~\eqref{eq:ZtH} (up to an overall normalization that cancels in ratios such as $Z_{\mathrm{tH}}/Z_{\mathrm{tH}}[0;0]$), without reweighting.

\section{Numerical results: confining phase and Witten effect}

\begin{figure}[t]
  \centering
  \includegraphics[width=0.8\textwidth]{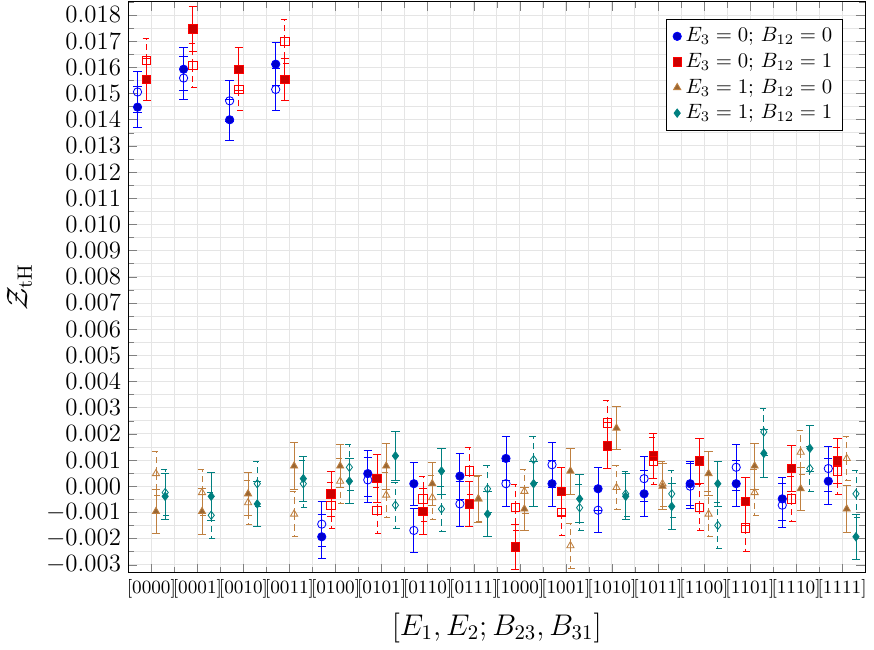}
  \caption{Confining-phase data for $Z_{\mathrm{tH}}[E;B]$ in $SU(2)$ YM on $T^4$.}
  \label{fig:conf}
\end{figure}

We focus on $SU(2)$ lattice YM on $T^4$ and measure $Z_{\mathrm{tH}}[E;B]$ for all combinations of electric fluxes $E_i\in\{0,1\}$ and magnetic fluxes $B_{ij}\in\{0,1\}$.
Figure~\ref{fig:conf} shows representative data in a confining regime.
$x$-axis enumerates $[E_1,E_2;B_{23},B_{31}]$ with $E_3$, $B_{12}$ fixed as in legend.
The filled symbols denote $Z_{\mathrm{tH}}$ itself, while the unfilled ones represent the conterparts of the duality equation.
The main qualitative observation is the ``light/heavy'' hierarchy:
\begin{equation}
  \frac{Z_{\mathrm{tH}}[E=0;B]}{Z_{\mathrm{tH}}[E=0;B=0]}\sim 1,
  \qquad
  \frac{Z_{\mathrm{tH}}[E\neq 0;B]}{Z_{\mathrm{tH}}[E=0;B=0]}\sim 0,
  \label{eq:lightheavy}
\end{equation}
consistent with confinement where nontrivial electric flux costs an area-law free energy.

\begin{figure}[t]
  \centering
  \includegraphics[width=0.8\textwidth]{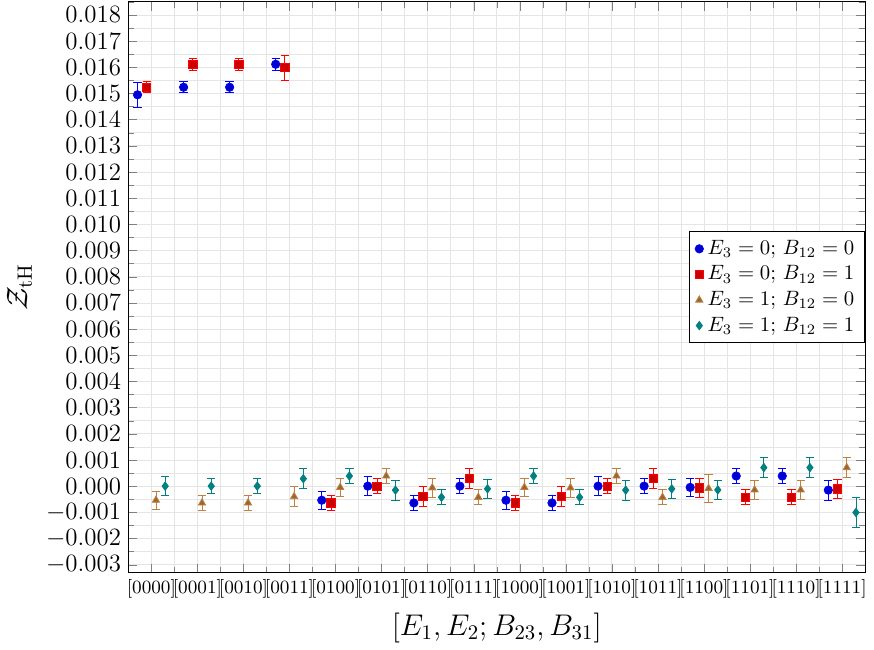}
  \caption{Averaged $Z_{\mathrm{tH}}[E;B]$ by Euclidean rotations.}
  \label{fig:average}
\end{figure}

Figure~\ref{fig:average} summarizes the averaged data and the observed pattern with less numerical errors.
Unfortunately, while the asymptotic behavior in the large volume limit is obtained by
\begin{equation}
  \frac{Z_{\mathrm{tH}}[E=0;B]}{Z_{\mathrm{tH}}[E=0;B=0]} - 1=O(e^{-\sigma L})
  \qquad
  \frac{Z_{\mathrm{tH}}[E\neq 0;B]}{Z_{\mathrm{tH}}[E=0;B=0]}=O(e^{-\sigma L}),
  \label{eq:tension}
\end{equation}
where $\sigma$ is the (dual) string tension,
it would be prohibitively expensive to determine $\sigma$ with the present statistics.

\begin{figure}[t]
  \centering
  \includegraphics[width=0.8\textwidth]{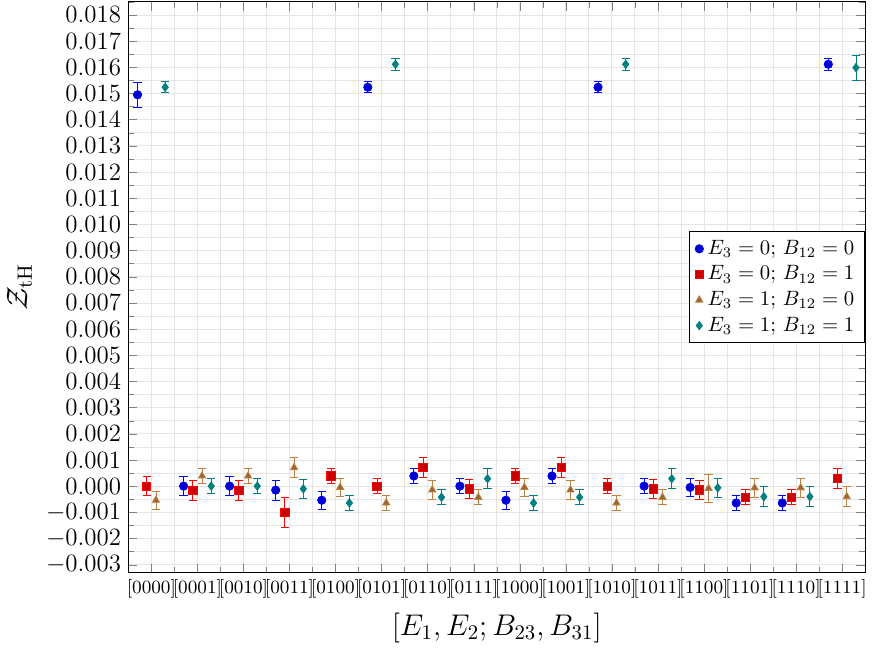}
  \caption{Shift relation \eqref{eq:witten} and oblique confinement (Witten effect).}
  \label{fig:witten}
\end{figure}

A further diagnostic is the Witten effect at $\theta=2\pi$, which implies a shift of electric flux by magnetic flux.
The 't~Hooft partition function with the $\theta$-term is given by Eq.~\eqref{eq:ZtH} by replacing $Z[B]$ with~$Z_\theta[B]$.
In our normalization this appears as
\begin{equation}
  Z_{\mathrm{tH},\theta=2\pi}[E_i;B_{jk}] = Z_{\mathrm{tH},\theta=0}[E_i + B_{jk}, B_{jk}],
  \label{eq:witten}
\end{equation}
illustrating oblique confinement as in Fig.~\ref{fig:witten}: the $\theta$ term effectively transmutes the ``light'' sector.

\section{Finite temperature: preliminary study and outlook}

\begin{figure}[t]
  \centering
  \includegraphics[width=0.98\textwidth]{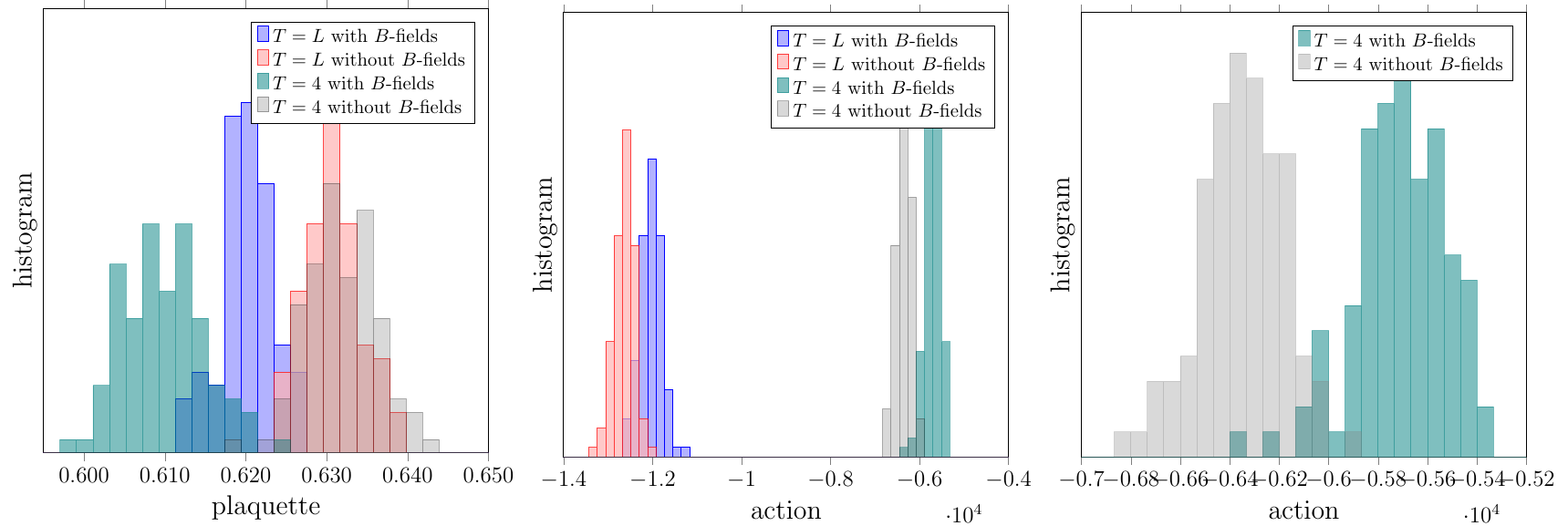}
  \caption{Distributions of plaquette and action. $L^3\times T$ with $L=8$, $\beta=2.4$, $B_{34}\in\{0,1\}$. The blue histogram denotes those of $T=L$, $B_{34}=1$, the red one is $T=L$, $B_{34}=0$ (zero-temperature); the green one is $T=L/2$, $B_{34}=1$, and the gray one is $T=L/2$, $B_{34}=0$ (finite-temperature).}
  \label{fig:boltz}
\end{figure}

For thermal geometries $T^3\times S^1$ near the deconfinement transition (small $S^1$), flux-sector physics becomes more delicate.
So far, we assumed that in our algorithm the total partition function for~$SU(N)/\mathbb{Z}_N$ can be decomposed by
\begin{equation}
 Z_{SU(N)/\mathbb{Z}_N} = \sum_B Z[B],\qquad N_{\mathrm{total}} = \sum_B N_{\mathrm{conf}}(B).
\end{equation}
Is $Z[B]$ truly a partition function?
If so, configurations in $Z[B]$ should be sufficiently thermalized in the conventional HMC for fixed~$B$.
When $P_F(B\to B')$ generates a different $B_{i4}'\neq B_{i4}$, a thermalized configuration $U'$ is far from the initial configuration $U$ in $Z[B]$.
Figure~\ref{fig:boltz} shows the distributions of values of plaquette and action for zero- and finite-temperature, illustrating the corresponding Boltzmann distribution.
Only $B_{34}$ can be taken as $0$ or $1$, otherwise $B_{\mu\nu}=0$.
For zero-temperature, the distributions for $B_{34}=0$, $1$ overlap each other, but for finite-temperature (when $S^1$ becomes smaller), those move away and stay farther apart.

An important practical question is whether $Z[B]$ computed by counting is robust against different thermalization times for different $B$ sectors.
This is tied to the ``separability'' of $\sum_B Z[B]$ and to constructing an efficient proposal kernel $P_F(B\to B')$ that respects sector-dependent thermalization scales.

\noindent
\textbf{Diagnosing (non-)thermalization and sector-dependent bias at finite temperature.}
Near the deconfinement transition on $T^3\times S^1$, the Markov chain may sample different flux sectors with substantially different relaxation times.
In that situation, the naive counting estimator in Eq.~\eqref{eq:counting} can be biased if measurements are taken before the within-sector dynamics has equilibrated after a change of $B$.
To make this issue quantitative, we monitor standard observables such as the plaquette and the gauge action, and compare their distributions in each sector with those obtained from a conventional HMC run at fixed $B$ (matched at the same bare parameters and geometry).
Figure~\ref{fig:boltz} illustrates these diagnostics by comparing the plaquette and action distributions between flux sectors at zero and finite temperature.
As a compact diagnostic, we also track the integrated autocorrelation time $\tau_{\mathrm{int}}$ of the plaquette within each sector, and the acceptance rate of the composite move $(U,B)\to(U',B')$ as functions of $(B,\beta,L,T)$.

\noindent
\textbf{A practical measurement protocol.}
In the thermal runs, after every accepted update that changes the temporal flux components $B_{i4}$, we discard $N_{\mathrm{therm}}$ subsequent trajectories (``sector re-thermalization'') before recording measurements.
Equivalently, we may record measurements only when the chain has remained in the same sector for at least $N_{\mathrm{therm}}$ trajectories.
We then verify stability by checking that the sector-resolved plaquette/action histograms agree (within statistics) between the first and second halves of the Monte Carlo history, and that the inferred $\widehat Z[B]$ is insensitive to moderate variations of $N_{\mathrm{therm}}$.
These checks provide a direct, observable-based criterion for whether sector counting can be interpreted as a reliable proxy for relative weights at finite temperature.

\noindent
\textbf{Tuning the flux proposal.}
Finally, the proposal kernel $P_F(B\to B')$ should be tuned so that sector changes are neither too rare (leading to poor exploration) nor too frequent (leading to persistent re-thermalization transients).
In practice, we adjust the proposal width (e.g.\ the parameter $\alpha$ in the Gaussian proposal used below) to obtain a stable acceptance rate and manageable $\tau_{\mathrm{int}}$, and we report these algorithmic diagnostics together with the physics observables.

As a first test, a simplified setup is that only $B_{34}=z\in\{0,1\}$ is turned on and employed a Gaussian proposal $P_F\propto e^{-\alpha(z-z')^2}$. We should find an empirical thermalization scale of MD time units to tune the parameter $\alpha$. Notably, the thermalization time/acceptance ratio in MD starting from $U[B_{34}=0]$ to $U[B_{34}=1]$ is quite different from that from $U[B_{34}=1]$ to $U[B_{34}=0]$.
The practical transition probability~$P_F$ in MD time units is larger than the longest thermalization time for each lattice parameter.

Based on our experience, the observation of $Z_{\mathrm{tH}}$ is hindered by too large systematic error of order~$Z_{\mathrm{tH}}[0;0]/3$, while statistical errors are comparable with those in Fig.~\ref{fig:average} .

Improving the finite-temperature analysis will require (i) larger statistics for finite-size scaling, (ii) systematic control of thermalization in each flux sector, and (iii) optimized $P_F$ kernels tailored to the sector structure.
It will also be interesting to compare our lattice estimates of $Z[B]$ and $Z_{\mathrm{tH}}$ with continuum predictions from symmetry-TFT/anomaly-inflow descriptions of center symmetry, in particular the phase factors and selection rules of twisted partition functions.
 
\section*{Acknowledgments}
This work was partially supported by Japan Society for the Promotion of Science (JSPS)
Grant-in-Aid for Scientific Research Grant Number
JP25K17402 (O.M.) and JP23K03418 (H.S.).
The numerical computations in this paper were carried out on Genkai, a
supercomputer system of the Research Institute for Information Technology
(RIIT), Kyushu University.
Our numerical codes, which can be found in
\url{https://github.com/o-morikawa/Gaugefields.jl},
is an extended version of \texttt{Gaugefields.jl} in JuliaQCD project~\cite{Nagai:2024yaf}.
O.M.\ acknowledges the RIKEN Special Postdoctoral Researcher Program
and RIKEN FY2025 Incentive Research Projects.

\bibliographystyle{utphys}
\bibliography{ref}
\end{document}